\documentclass[aps,pra,reprint,superscriptaddress,showpacs,showkeys,floatfix]{revtex4-1}

\usepackage{graphicx}
\usepackage{color}
\usepackage{amssymb}
\usepackage{amsmath}
\usepackage{epsfig}
\usepackage{bm}
\usepackage[american]{babel}
\usepackage{hyperref}
\usepackage{braket}

\usepackage[T1]{fontenc} 
\usepackage{txfonts}

\DeclareGraphicsExtensions{.pdf,.png,.jpg}

\hypersetup{colorlinks=true,urlcolor=blue,breaklinks=true,citecolor=blue,linkcolor=blue,pdfstartview=FitH,pdfpagemode=UseNone}

\newcommand{\inda}{a}
\newcommand{\indb}{b}
\newcommand{\indc}{c}

\newcommand{\ve}[1]{\mathbf{#1}}                 
\newcommand{\spin}{s}                                       

\newcommand{\note}{\color{black}}

\begin{document}

\title{Creation of a Dirac monopole--antimonopole pair in a spin-1 Bose--Einstein condensate}
\date{\today}

\author{Konstantin Tiurev} \email{konstantin.tiurev@gmail.com}
\affiliation{QCD Labs, QTF Centre of Excellence, Department of Applied Physics, Aalto University, P.O. Box 13500, FI-00076 Aalto, Finland}
\author{Pekko Kuopanportti}
\affiliation{Department of Physics, University of Helsinki, P.O. Box 43, FI-00014 Helsinki, Finland}
\author{Mikko M{\"o}tt{\"o}nen}
\affiliation{QCD Labs, QTF Centre of Excellence, Department of Applied Physics, Aalto University, P.O. Box 13500, FI-00076 Aalto, Finland}

\begin{abstract}
We theoretically demonstrate that a pair of Dirac monopoles with opposite synthetic charges can be created within a single spin-1 Bose--Einstein condensate by steering the spin degrees of freedom by external magnetic fields. Although the net synthetic magnetic charge of this configuration vanishes, both the monopole and the antimonopole are accompanied by vortex filaments carrying opposite angular momenta. Such a Dirac dipole can be realized experimentally by imprinting a spin texture with a nonlinear magnetic field generated by a pair of coils in a modified Helmholtz configuration. We also investigate the case where the initial state for the dipole-creation procedure is pierced by a quantized vortex line with a winding number $\kappa$. It is shown that if $\kappa = -1$, the resulting monopole and antimonopole lie along the core of a singly quantized vortex whose sign is reversed at the locations of the monopoles. For $\kappa = -2$, the monopole and antimonopole are connected by a vortex line segment carrying two quanta of angular momentum, and hence the dipole as a whole is an isolated configuration. In addition, we simulate the long-time evolution of the dipoles in the magnetic field used to create them. For $\kappa=0$, each of the semi-infinite doubly quantized vortices splits into two singly quantized vortices, as in the case of a single Dirac monopole. For $\kappa = -1$ and $\kappa = -2$, the initial vortices deform into a vortex with a kink and a vortex ring, respectively.
\end{abstract}
\maketitle

\section{Introduction}\label{introduction}
One of the open questions in particle physics is whether or not magnetic monopoles exist~\cite{doi:10.1146/annurev.ns.34.120184.002333}. These hypothetical particles play significant roles in nature, and their existence would have far-reaching implications for both, quantum and classical theories of electromagnetism. Dirac's theory of monopoles consistent with quantum mechanics and the gauge invariance of the electromagnetic field sparked the search for these elusive particles almost a century ago~\cite{Dir1931.PRSLA133.60}. Dirac considered a charged quantum-mechanical particle in the presence of a static magnetic field of a monopole. He showed that the wave function of the scalar particle is inevitably accompanied by a semi-infinite line of vanishing density, which is often referred to as a Dirac nodal line. The nodal line is a physically observable feature distinct from the Dirac string, which, in contrast, is a gauge-dependent object in the accompanying vector potential. Another major step in the monopole problem was taken in 1974 independently by 't~Hooft~\cite{HOOFT1974276} and Polyakov~\cite{Polyakov.1974}, who showed the natural appearance of monopole solutions in grand unified theories~\cite{PhysRevLett.32.438}. 

While monopoles in the natural magnetic field remain elusive to experimental observation, a great deal of effort has been put into the search for corresponding configurations in experimentally more tractable systems. Analogs of magnetic monopoles have been found in a number of classical systems such as exotic spin ices~\cite{Castelnovo:2008aa,Morris411} and nematic liquid crystals~\cite{PhysRevE.93.062703}. Creation of a Dirac monopole in a ferromagnetic Bose--Einstein condensate (BEC) was recently achieved experimentally, providing the first known realization of Dirac's monopole theory~\cite{Ray2014.Nat505.657,Pie2009.PRL103.030401,Sav2003.PRA68.043604}. An isolated monopole similar to the 't Hooft--Polyakov monopole~\cite{Sto2001.PRL87.120407} was consequently created in a quantum-mechanical order parameter of the polar-phase spin-1 BEC~\cite{Ray2015.Sci348.544}. In Refs.~\cite{Tiurev2015.pra.93.033638,Tiurev2016.pra.94.053616}, the detailed evolution of the isolated polar-phase monopoles was studied numerically under conditions similar to their experimental realization. Driven by the phase transition from the polar state to the natural ferromagnetic ground state of the condensate~\cite{Sad2006.Nat443.312}, the isolated monopole was predicted to relax spontaneously into a polar-core vortex~\cite{Tiurev2016.pra.94.053616} or a ground-state Dirac monopole configuration~\cite{Tiurev2015.pra.93.033638,Ruo2011.PRA84.063627} depending, respectively, on whether the quadrupole magnetic field was absent or present during the evolution. The latter prediction was subsequently verified experimentally by Ollikainen \emph{et al.}~\cite{PhysRevX.7.021023}.

The aforementioned experiments~\cite{Ray2014.Nat505.657,Ray2015.Sci348.544,PhysRevX.7.021023} utilized a precise control of the atomic spins by steering them with a combination of uniform and quadrupole magnetic fields, the latter generated by anti-Helmholtz current coils. In this article, we propose modifying the monopole-creation method implemented in Refs.~\cite{Ray2015.Sci348.544,Ray2014.Nat505.657,PhysRevX.7.021023} by replacing the anti-Helmholtz coils with larger coils carrying parallel currents~\cite{PhysRevA.35.1535}. We show theoretically that this adjustment makes it possible to imprint a pair of synthetic Dirac monopoles within a single ferromagnetic condensate, each carrying opposite magnetic charge and attached to a separate nodal line. Creation of such monopole--antimonopole pairs in the natural magnetic field has been discussed in the context of high-energy experiments, e.g., via electron--positron~\cite{MUSSET1983333,PhysRevD.29.1524,PINFOLD1993407}, proton--antiproton~\cite{PhysRevLett.85.5292}, and proton--positron collisions~\cite{TheH1Collaboration2005} in accelerators. Experiments with nematic liquid crystals have shown the spontaneous appearance of monopole--antimonopole pairs resulting from unwinding of textures~\cite{CHUANG1336} via the Kibble--Zurek mechanism~\cite{Zurek:1985aa,0305-4470-9-8-029}. Thus, it is interesting to investigate the behavior of monopole pairs in the context of BECs, especially the dynamics of interconnected monopoles. An analytical solution for such dipoles was constructed in Ref.~\cite{Sav2003.PRA68.043604}, but no experimental scheme for their creation has been proposed so far.

We simulate both the creation process and the subsequent evolution of the monopole--antimonopole pairs for parameters pertaining to $^{87}$Rb atoms. Our simulations agree well with analytical expressions for monopole--antimonopole solutions. We also show that  doubly quantized vortices split into pairs of singly quantized vortices for energetic reasons~\cite{Kaw2004.PRA70.043610}, in analogy with the case of a single Dirac monopole~\cite{PhysRevX.7.021023,Ruo2011.PRA84.063627}. We furthermore investigate two additional cases: (a) the monopoles are located at the core of a single-quantum vortex whose vorticity is effectively reversed at the locations of the monopoles, and (b) the monopole and antimonopole are connected by a doubly quantized vortex line segment. The decay dynamics are also studied in both cases and shown to result in nontrivial vortex configurations, such as polar-core vortex ring~\cite{PhysRevLett.86.3934}.

This paper is organized as follows. Section~\ref{theory} provides a brief theoretical introduction to the synthetic electromagnetism in spinor condensates. {\note In Sec.~\ref{setup}, we describe our proposed experimental protocol and identify expected technical challenges in its realization.}
We discuss the analytical solutions for Dirac dipoles in Sec.~\ref{analytics}. The details of numerical simulations are given in Sec.~\ref{methods}. The main results of this paper are presented and discussed in Sec.~\ref{results}, and the conclusions are given in Sec.~\ref{conclusions}. 

\section{Theoretical background}\label{theory}

In the context of spin-1 condensates, a Dirac monopole configuration is revealed by writing the mean-field order parameter of the condensate as $\Psi = \psi \zeta$, where $\psi = \sqrt{n}e^{i\varphi}$ is the scalar part of the order parameter, $n = \Psi^{\dagger}\Psi$, and $\zeta = (\zeta_{1},\zeta_{0},\zeta_{-1})^{\textrm{T}}$ is a normalized spinor expressed in the $z$-quantized basis $\{\ket{F=1, m_z=1},\ket{F=1, m_z=0},\ket{F=1, m_z=-1}\}$~\cite{Kaw2012.PhysRep520.253}. The artificial, or synthetic, radial magnetic field acting on the scalar part of the condensate order parameter arises from the synthetic vector and scalar gauge potentials~\cite{Lee:2018ab,Lin:2011aa,Kaw2012.PhysRep520.253}
\begin{equation}\label{eq:A}
\mathbf{A}^*(\mathbf{r},t) = i \zeta^{\dagger}(\mathbf{r},t) \nabla \zeta(\mathbf{r},t),
\end{equation}
and
\begin{equation}\label{eq:Phi}
\Phi^*(\mathbf{r},t) = i \zeta^{\dagger}(\mathbf{r},t) \frac{\partial}{\partial t} \zeta(\mathbf{r},t),
\end{equation}
respectively~\cite{note_asterisk}. Thus, the temporally and spatially varying spinor field $\zeta(\mathbf{r},t)$ gives rise to synthetic electromagnetism. In particular, the spinor imprinted in Ref.~\cite{Ray2014.Nat505.657} reads 
\begin{equation}\label{eq:Diracspinor}
\zeta(\theta, \phi) = 
\begin{pmatrix}
\frac{1}{2}(1 - \cos\theta) \\
\frac{1}{\sqrt{2}} e^{i\phi} \sin\theta \\
\frac{1}{2}e^{2i\phi}(1 + \cos\theta)
\end{pmatrix},
\end{equation}
where $\theta$ and $\phi$ are the polar and azimuthal angles, respectively. By substituting Eq.~\eqref{eq:Diracspinor} into Eq.~\eqref{eq:A} and calculating its curl, we obtain the synthetic magnetic field
\begin{equation}\label{eq:Dirac_monopole}
\mathbf{B}^*(\mathbf{r},t) = \hbar [ \nabla \times \mathbf{A}^*(\mathbf{r},t) ] = \hbar\frac{\mathbf{\hat{r}}}{r^2},
\end{equation}
where we have left out any singularity corresponding to the Dirac string since it can be fully removed by a proper gauge transformation~\cite{note:gauge}. Thus, the scalar part of the condensate and the synthetic magnetic field generated by the spinor field play, respectively, the roles of the charged scalar particle and the field of the classical magnetic monopole discussed in Dirac's seminal paper~\cite{Dir1931.PRSLA133.60}. 

The synthetic magnetic field is equal to the physically observable superfluid vorticity
\begin{equation}\label{eq:v}
\mathbf{\Omega}_{\textrm{s}}(\mathbf{r},t) = \nabla \times \mathbf{v}_s(\mathbf{r},t)
\end{equation}
almost everywhere, that is, away from the nodal lines. Here,
\begin{equation}\label{eq:v}
\mathbf{v}_{\textrm{s}}(\mathbf{r},t) = \frac{\hbar}{m}  \Big{[} \nabla \phi(\mathbf{r},t) - \mathbf{A}^*(\mathbf{r},t) \Big{]} 
\end{equation}
is the superfluid velocity and $m$ is the mass of the atoms. 

\begin{figure}[t]	
\includegraphics[width=0.95\columnwidth]{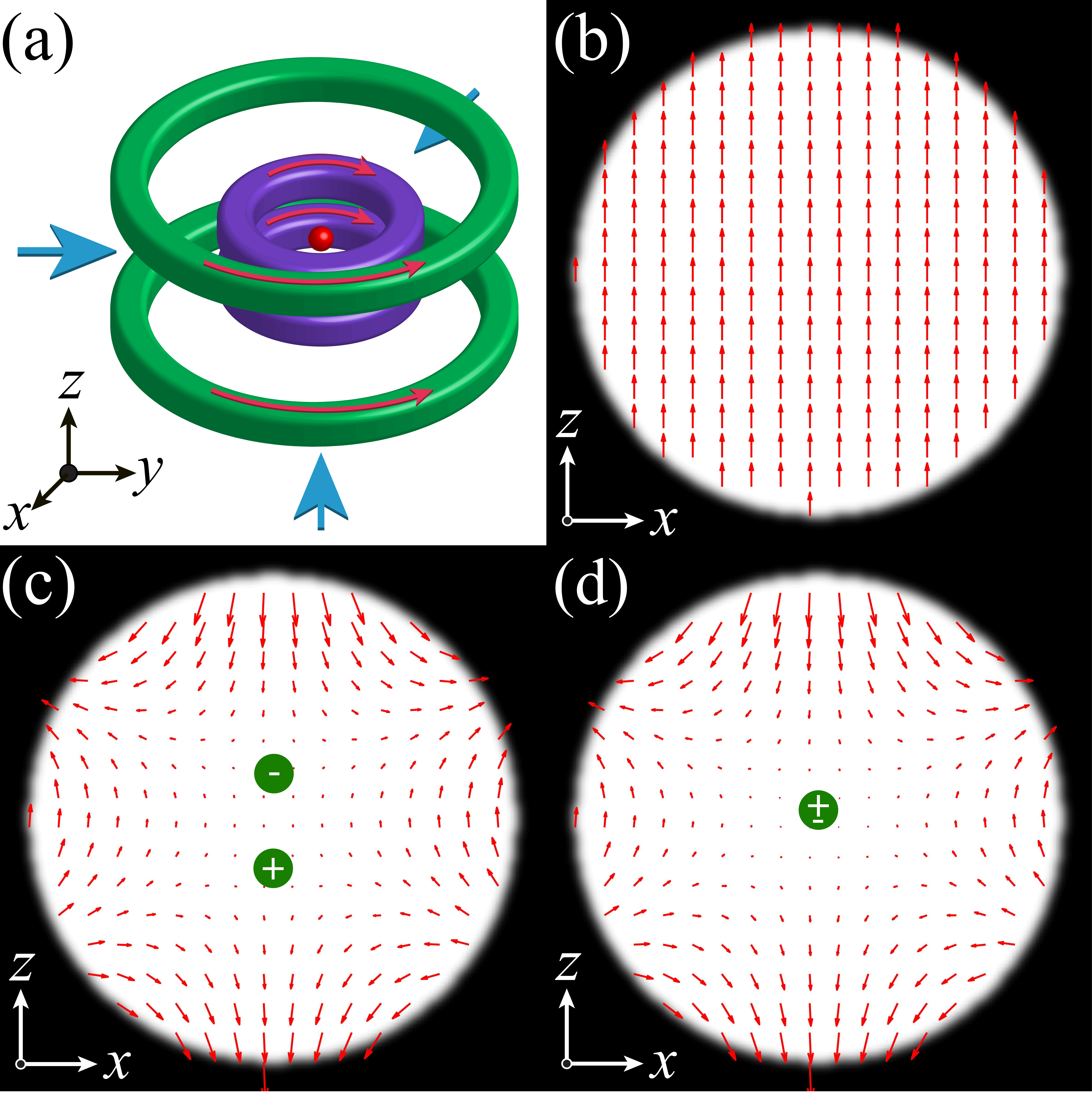}
\caption{\label{fig:1} Schematic representation of (a)~the experimental setup and (b)--(d)~the magnetic field it generates. (a)~A schematic showing the coils that produce the uniform~(larger, green loops) and non-uniform~(smaller, magenta loops) magnetic fields. Electric current, running in parallel in each pair of coils, is shown with red arrows. The condensate~(red sphere) is confined in an optical trap generated by three orthogonal laser fields~(large blue arrows). (b)--(d)~The axially symmetric magnetic field when (b)~the bias field is zero; (c) the field zeros are brought into different hemispheres of the condensate; (d) the field zeros are located at the center of the condensate. Dots with $+$ and $-$ in (c) show the locations of the monopole and antimonopole, respectively; the dot with $\pm$ in (d) indicates coincident locations. The field of view in (b)--(d) is 15.5$\times$15.5 $\mu$m$^2$.}
\end{figure}

\section{Experimental setup}\label{setup}
To facilitate the experimental creation of Dirac monopole--antimonopole pairs, we propose a formally simple change in the design of the magnetic coils compared to the previous method of creating a Dirac monopole~\cite{Ray2014.Nat505.657}. Specifically, we suggest replacing a pair of magnetic coils in the anti-Helmholtz configuration with coils carrying current in the same direction. Denoting the radii of the coils in the pair by $R_{1,2}$ and their axial~($z$) locations by $A_{1,2}$, the magnetic field generated by a pair of coils in cylindrical coordinates $(\,{\rho} = \sqrt{x^2+y^2},\phi = \mathrm{arctan2}(y,x),z\,)$ reads~\cite{PhysRevA.35.1535}
\begin{equation}\label{eq:B_tot}
\mathbf{B}_{\textrm{d}} = (B_{z_{1}} + B_{z_{2}})\mathbf{\hat{z}} + (B_{\rho_{1}}+B_{\rho_{2}}) \hat{\boldsymbol{\rho}}, 
\end{equation}
where
\begin{align}
B_{z_{1,2}}(\rho,z) &=  \frac{\mu N I}{2\pi}\frac{1}{\sqrt{(R_{1,2}+\rho)^2 + (z-A_{1,2})^2}} \Big{[} K(k_{1,2}^2) \nonumber
\\&+ \frac{R_{1,2}^2 -\rho^2 - (z-A_{1,2})^2}{(R_{1,2}-\rho)^2 + (z-A_{1,2})^2}E(k_{1,2}^2)  \Big{]} \label{eq:Bz} \\
\intertext{and}
B_{\rho_{_{1,2}}}(\rho,z) &=  \frac{\mu N I}{2\pi\rho}\frac{z-A_{1,2}}{\sqrt{(R_{1,2}+\rho)^2 + (z-A_{1,2})^2}} \Big{[}-K(k_{1,2}^2) \nonumber
\\ &+ \frac{R_{1,2}^2 + \rho^2 + (z-A_{1,2})^2}{(R_{1,2}-\rho)^2 + (z-A_{1,2})^2}E(k_{1,2}^2)  \Big{]} \label{eq:Br}
\end{align}
are the axial and radial components of the magnetic field due to each coil. Here $I$ is the current in the coil, $N$ is the number of windings, $\mu$ is the permeability, and $K(k^2)$ and $E(k^2)$ are the complete elliptic integrals with
\begin{equation}\label{eq:k12}
k_{1,2}^2 = \frac{4R_{1,2}\rho}{(R_{1,2}+\rho)^2 + (z-A_{1,2})^2}.
\end{equation}

In the modified Helmholtz coils~[smaller~(magenta) loops in Fig.~\ref{fig:1}(a)], we choose the radii of the coils to be larger than their separation in order to maximize the curvature of the magnetic field. Such a configuration generates a magnetic field $\mathbf{B}_{\textrm{d}}(\mathbf{r})$ with two zeros on the symmetry axis $z$, equidistant from the origin. We also use standard Helmholtz coils~[larger~(green) loops in Fig.~\ref{fig:1}(a)] to create a highly uniform magnetic field $B_{\textrm{b}}(t)$, which we refer to as the bias field and use to move the field zeros along the $z$ axis. 
The total magnetic field thus reads
\begin{equation}\label{eq:Btot}
\begin{aligned}
\mathbf{B}_{\textrm{tot}}(\mathbf{r},t) = \mathbf{B}_{\textrm{d}}(\mathbf{r}) +  B_{\textrm{b}}(t)\mathbf{\hat{z}}.
\end{aligned}
\end{equation}

{\note We assume that initially $B_{\textrm{b}}(0) =  -|\mathbf{B}_{\textrm{d}}(\boldsymbol{0})|  + \delta B$, which for a sufficiently strong offset field $\delta B > 0$ yields the almost uniform field configuration shown in Fig.~\ref{fig:1}(b), ; the application of such a field to the spin-1 BEC renders the atomic. Subsequently, $B_{\textrm{b}}(t)$ is adiabatically decreased, which moves the two field zeros towards the origin.
We refer to this step as the creation ramp.} Depending on the final value chosen for the bias field, the creation ramp either brings both magnetic field zeros to $(\rho,z)=(0,0)$, as shown in Fig.~\ref{fig:1}(d), or stops earlier, placing the two field zeros in different hemispheres of the condensate cloud, as in Fig.~\ref{fig:1}(c).

{\note We now turn to the question of the technical feasibility of the proposed scheme in real experiments. 
The anti-Helmholtz coils used in Refs.~\cite{Ray2014.Nat505.657,Ray2015.Sci348.544} generate magnetic field gradients strong enough to steer the atomic spins even when modest electric currents are used.  
In contrast, the magnetic field gradient generated in the proposed scheme is orders of magnitudes weaker for the same current values. 
To achieve a sufficiently strong gradient, we choose the parameters of the magnetic coils in Eqs.~\eqref{eq:Bz} and \eqref{eq:Br} to be $R_1=R_2=4$~cm, $A_1 = -A_2 = 1$~cm, $I=400$~A, and $N=500$.
Two technical challenges arise immediately. First, the total electric current of 200~kA through 4~cm coil is extremely high, and we are not aware of any experimental realizations of resistive electromagnetic coils with such parameters.
Second, in order to cancel the uniform component of the magnetic field, which for the chosen parameters exceeds 3~T, the electric currents through both pairs of coils in Fig.~\ref{fig:1}(a) must be controlled with unprecedented precision.
We suggest superconducting coils as a possible solution to these challenges. 
Superconducting setups that generate suitable magnetic fields exist~\cite{0953-2048-26-10-105014,0953-2048-18-2-026} and may be applicable in future experiments. 
For this theoretical proposal, however, we will proceed with the parameters chosen above and leave the technical realization as an open question for future investigation.}

\section{Analytical solution}\label{analytics}

During the adiabatic creation ramp, the spins rotate by an angle \mbox{$\beta=\textrm{arccos}(\mathbf{B_{\textrm{tot}}}\cdot \mathbf{\hat{z}}/|\mathbf{B_{\textrm{tot}}}|)$} about the azimuthally varying axis $\mathbf{n} = -\mathbf{\hat{x}}\sin\phi + \mathbf{\hat{y}}\cos\phi$, where $\mathbf{B}_{\textrm{tot}}$ is defined in Eq.~\eqref{eq:Btot}. In the $z$-quantized basis, this spin rotation corresponds to the transformation matrix
\begin{equation}
\begin{aligned}\label{eq:rotationR}
\mathcal{R}(\beta, \phi) = \exp \Big{(}  -i \frac{\mathbf{F} \cdot \mathbf{\hat{n}}}{\hbar} \beta \Big{)},
\end{aligned}
\end{equation}
where $\mathbf{F} = (F_x, F_y, F_z)$ is a vector of the dimensionless spin-1 matrices satisfying $\left[F_\inda,F_\indb\right]=i\varepsilon_{\inda\indb\indc}F_\indc$, $\varepsilon_{\inda\indb\indc}$ is the Levi-Civita symbol, and $\inda,\indb,\indc \in \left\{x,y,z\right\}$. Applied to the initial-state spinor $\zeta = (1,0,0)^{\textrm{T}}$, this rotation yields
\begin{equation}\label{eq:spinor}
\begin{aligned}
\zeta_0(\beta, \phi) = 
\mathcal{R}(\beta, \phi)
\begin{pmatrix}
1 \\
0 \\
0
\end{pmatrix}
 = 
\begin{pmatrix}
\frac{1}{2}(1 + \cos\beta) \\
\frac{1}{\sqrt{2}} e^{i{\phi}} \sin\beta \\
\frac{1}{2}e^{2i{\phi}}(1 - \cos\beta)
\end{pmatrix}.
\end{aligned}
\end{equation}

We first present the analytical solution for the case of overlapping monopoles~[Fig.~\ref{fig:1}(d)], as it allows for a simple expression of the spinor and the corresponding synthetic magnetic field, similar to Eqs.~\eqref{eq:Diracspinor} and \eqref{eq:Dirac_monopole} for a single Dirac monopole. When the two field zeros are brought all the way to the center $(\rho,z)=(0,0)$, the adiabatic creation ramp rotates the spin states by an angle $\beta(\mathbf{r}) = \pi - 2\theta$, turning Eq.~\eqref{eq:spinor} into
\begin{equation}\label{eq:spinor_center}
\begin{aligned}
\zeta_{\textrm{c}}(\theta, \phi)
= 
\begin{pmatrix}
\sin^2{\theta}\\
\sqrt{2}e^{i \phi}\sin{\theta}\cos{\theta}\\
e^{2 i \phi}\cos^2{\theta}
\end{pmatrix}.
\end{aligned}
\end{equation}
Insertion of the spinor of Eq.~\eqref{eq:spinor_center} into Eq.~\eqref{eq:A} and calculation of the synthetic magnetic field as $\mathbf{B}^* = \hbar \nabla \times \mathbf{A}^*$ yields 
\begin{eqnarray}\label{eq:Dipole_A}
\mathbf{A_{\textrm{c}}^*}(\mathbf{r}) = -\frac{2 \cos{\theta} \cot{\theta}}{|r|} \hat{\pmb{\phi}},
\end{eqnarray}
and
\begin{eqnarray}
\mathbf{B_{\textrm{c}}^*}(\mathbf{r}) = \hbar\frac{4\cos{\theta}}{|r|^2} \hat{\mathbf{r}}.
\end{eqnarray}
Thus, a pointlike dipole creates a radial synthetic magnetic field that changes its sign at $z=0$. The vector potential $\mathbf{A_{\textrm{c}}^*}$ contains two singularities that lie on the positive and negative $z$ axes and coincide with vortex cores in the $\ket{-1}$ component of the condensate. The singularities in the vector potential~\eqref{eq:Dipole_A}  are not physically observable and can be removed by applying an appropriate gauge transformation, as it was shown in Ref.~\cite{note:gauge} for a Dirac monopole.
\begin{figure}[t]	
\includegraphics[width=0.85\columnwidth]{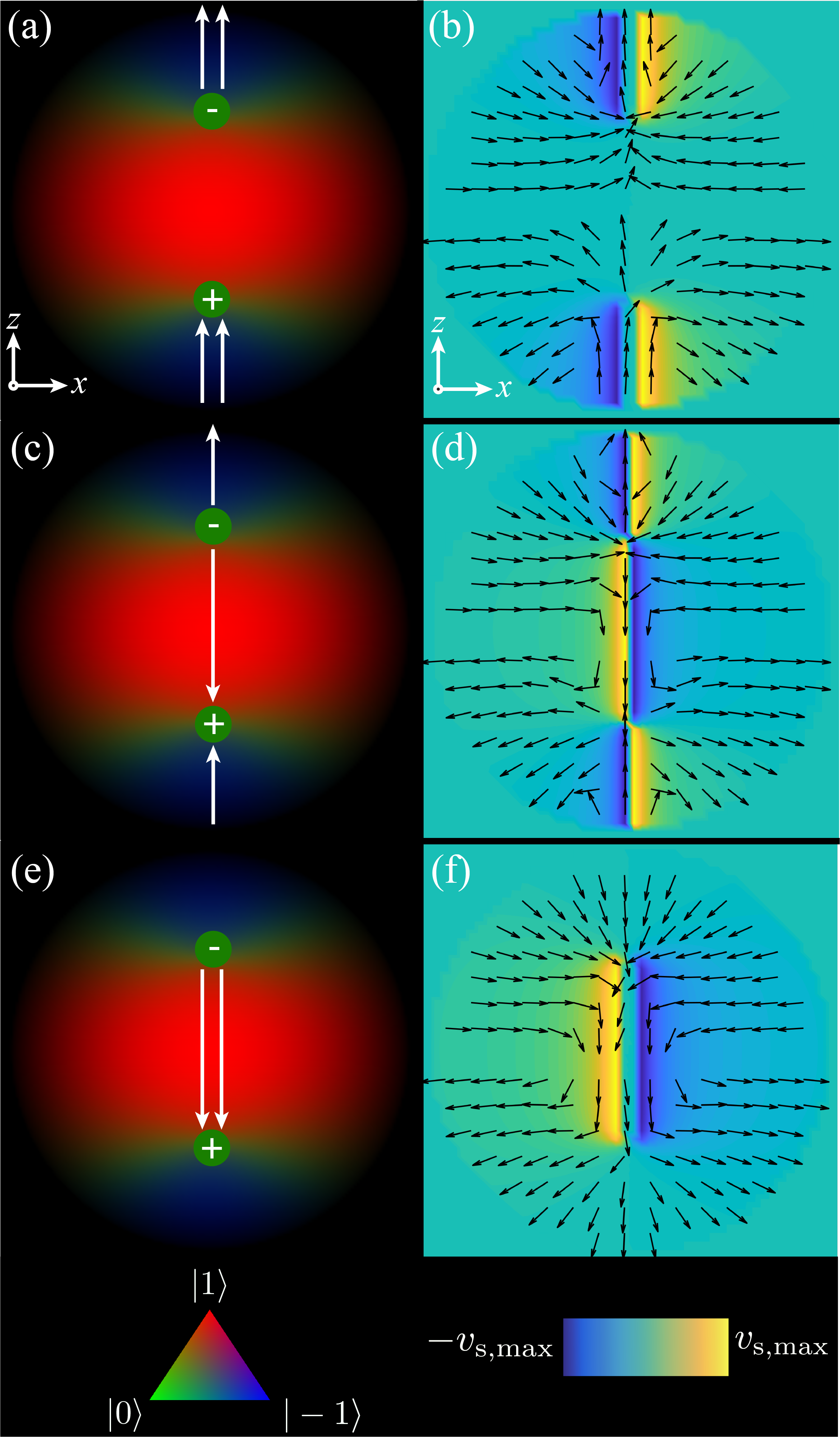}
\caption{\label{fig:2}{Dirac dipoles obtained from analytical solutions.} (a,c,e)~The $y$-integrated color-composite density of a Dirac dipole pierced by a vortex with a winding number (a)~$\kappa=0$, (b)~$\kappa=-1$, and (c)~$\kappa=-2$. Green dots and white arrows schematically represent the monopoles and the attached vortices. (b,d,f) The direction of the synthetic magnetic field in the $xz$ plane with the $y$ component of the superfluid velocity shown by the background. The field of view is 15.5$\times$15.5 $\mu$m$^2$.}
\end{figure}
\begin{figure}[t]	
\includegraphics[width=0.85\columnwidth]{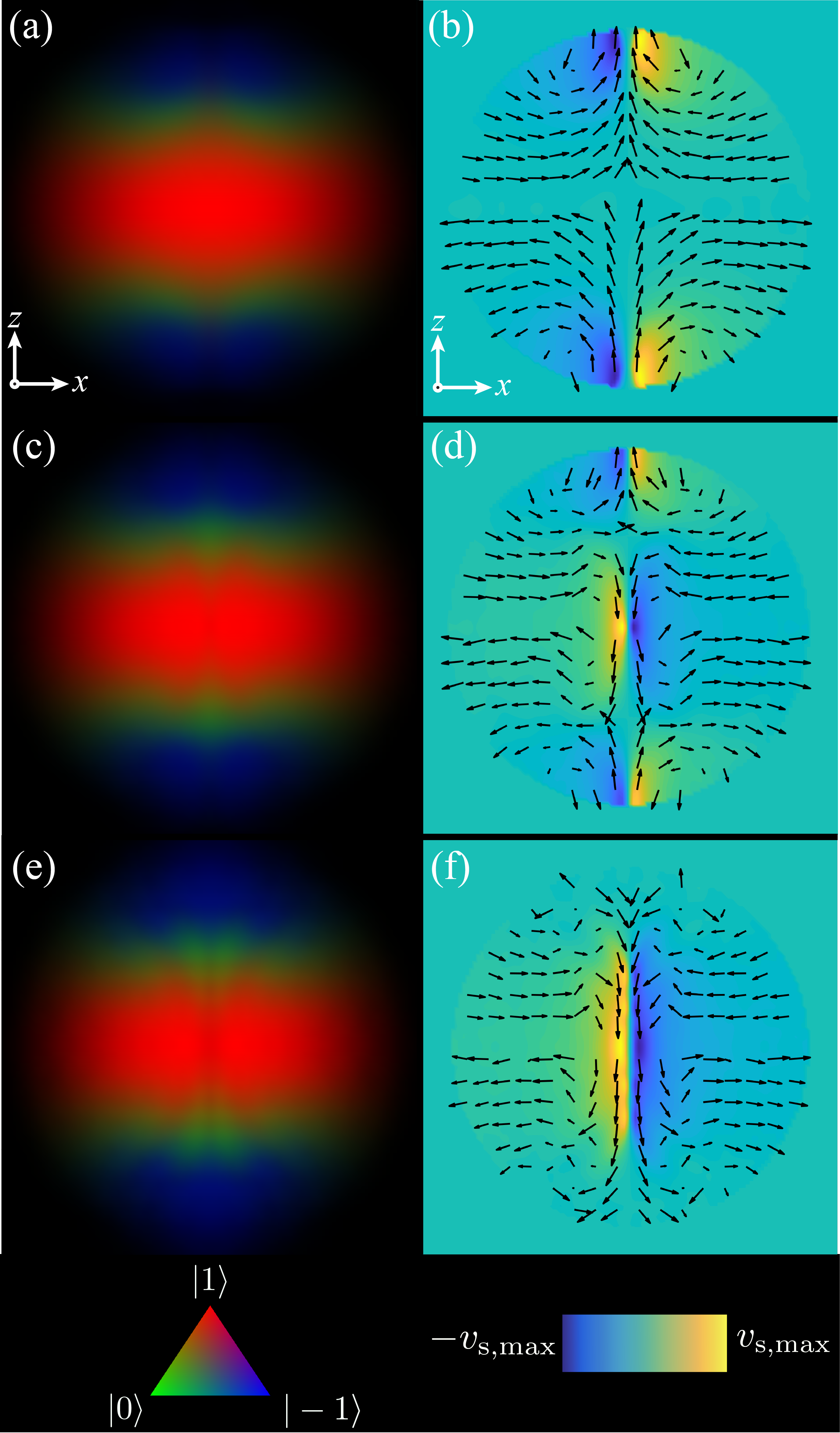}
\caption{\label{fig:3}{Dirac dipole configurations obtained from numerical simulations.} (a,c,e)~The $y$-integrated color-composite density of a Dirac dipole initially pierced by a vortex with a winding number (a)~$\kappa=0$, (b)~$\kappa=-1$, and (c)~$\kappa=-2$. (b,d,f) The direction of the synthetic magnetic field in the $xz$ plane with the $y$ component of the superfluid velocity shown by the background. The field of view is 15.5$\times$15.5 $\mu$m$^2$.}
\end{figure}
{However, as we will discuss in more detail in Sec.~\ref{results}, the monopole and antimonopole will, in practice, remain separated even if the field zeros are taken to the center of the condensate.} The more realistic configuration thus corresponds to the case shown in Fig.~\ref{fig:1}(c), where the field zeros are displaced from the condensate center. The corresponding spinor fields are presented in Figs.~\ref{fig:2}(a) and~\ref{fig:2}(b). The points where the external magnetic field vanishes are also the centers of the synthetic magnetic monopoles with opposite charges 
\begin{equation}\label{eq:monopole_charge}
\mathcal{Q} = \frac{1}{8\pi} \int_{\Sigma_{\pm 1}} \textrm{d}^2 \sigma_i \varepsilon_{ijk} \varepsilon_{abc} \hat{\mathbf{B}}^*_a
\partial_j \hat{\mathbf{B}}^*_b
\partial_k \hat{\mathbf{B}}^*_c 
= \pm 1,
\end{equation}
where $\Sigma_{+1}$ and $\Sigma_{-1}$ are surfaces enclosing the field zeros in the northern and southern hemispheres, respectively, and $\hat{\mathbf{B}}^* = \mathbf{B}^*/|\mathbf{B}^*|$ is the unit vector of the synthetic magnetic field. Each monopole acts as the termination point for a doubly quantized vortex line penetrating from outside into the condensate.

A creation procedure identical to that described above but applied to a condensate initially pierced by an axial vortex with a winding number $\kappa$ generates the spinor
\begin{equation}\label{eq:spinor_kappa}
\begin{aligned}
\zeta_{\kappa}(\beta, \phi) = 
\mathcal{R}(\beta, \phi)
\begin{pmatrix}
e^{i\kappa\phi} \\
0 \\
0
\end{pmatrix}
= 
e^{i\kappa\phi}\zeta_0.
\end{aligned}
\end{equation}
We show the spinor configurations of Eq.~\eqref{eq:spinor_kappa} for $\kappa = -1$ and $\kappa = -2$ in Figs.~\ref{fig:2}(c)--\ref{fig:2}(f). The $\kappa = -1$ dipole in Figs.~\ref{fig:2}(c) and \ref{fig:2}(d) lies at the core of a singly quantized vortex that reverses its vorticity at the locations of the monopoles. For the case $\kappa=-2$ shown in Figs.~\ref{fig:2}(e) and \ref{fig:2}(f), the initially imprinted vortex extending outside the condensate is cancelled by a Berry phase~\cite{Ber1984.ProcRSocA392.45} accumulated during the creation ramp. The resulting monopole and antimonopole are connected by a vortex with free ends, and the dipole as a whole is an isolated configuration similar to the dipole solution constructed in  Ref.~\cite{Sav2003.PRA68.043604}.

\section{Simulation methods}\label{methods}
Nonadiabatic effects arising from the harmonic potential, kinetic energy, and interatomic interactions cause the dipole state to deviate from the ideal case described by Eqs.~\eqref{eq:spinor} and \eqref{eq:spinor_kappa}. In order to take these effects into account, we simulate the creation process using the mean-field Gross--Pitaevskii~(GP) equation~\cite{Ho1998.PRL81.742,Ohm1998.JPSJ67.1822}
\begin{equation}
\begin{aligned}\label{eq:GP}
i\hbar \frac{\partial}{\partial t}{\Psi}(\mathbf{r},t) &= \Big[ -\frac{\hbar^2}{2m}\nabla^2 + V(\mathbf{r}) + g_{F}\mu_{\textrm{B}} \mathbf{B}_{\textrm{tot}}(\mathbf{r},t) \cdot \mathbf{F}  
\\& +  g_\mathrm{d} n(\mathbf{r},t) + g_\mathrm{s} n(\mathbf{r},t) \mathbf{\spin}(\mathbf{r},t)\cdot \mathbf{F} \Big]{\Psi}(\mathbf{r},t),
\end{aligned}
\end{equation}
where $g_F$ is the Land\'e factor, $\mu_{\textrm{B}}$ is the Bohr magneton, $\ve{\spin}={\zeta}^\dagger \mathbf{F} {\zeta}$ is the local average spin and $V(\mathbf{r}) = m \omega^2r^2/2$ is an external optical trapping potential which we assume to be spherically symmetric and harmonic with frequency $\omega$. The density--density and spin--spin couplings are defined by $g_\mathrm{d}=4\pi\hbar^2(a_0+2a_2)/3m$ and $g_\mathrm{s}=4\pi\hbar^2(a_2-a_0)/3m$, respectively, where $a_f$ is the $s$-wave scattering length corresponding to the scattering channel with total two-atom hyperfine spin $f$~\cite{Kaw2012.PhysRep520.253}.

\begin{figure}[t]	
\includegraphics[width=0.85\columnwidth]{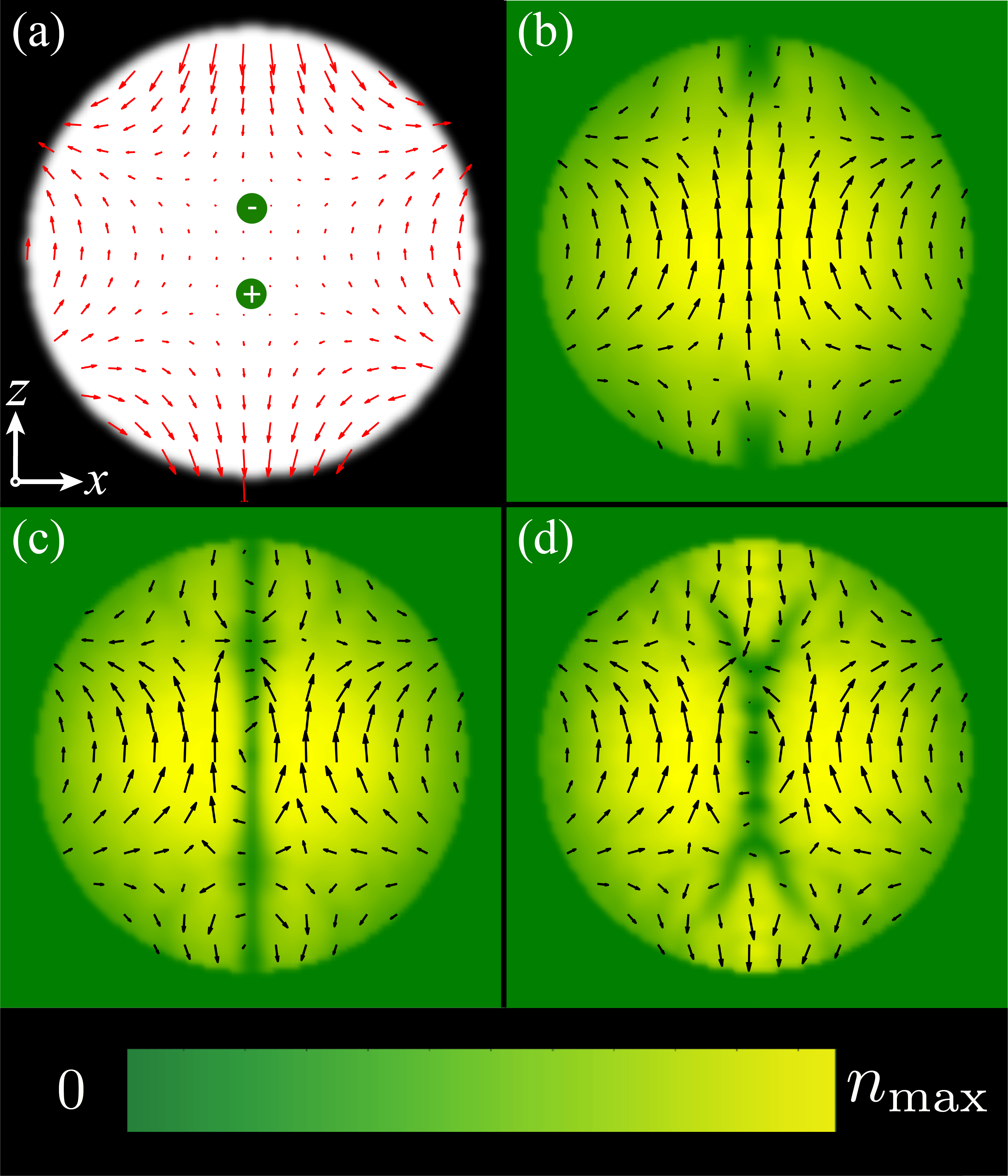}
\caption{\label{fig:4}{Spin fields of Dirac dipoles.} (a)~The magnetic field and (b)--(d)~spin fields right after the creation ramp. Panels (b), (c), and (d) correspond to a Dirac dipole imprinted into an initial state with a vortex of charge $\kappa = 0$, $\kappa = -1$, and $\kappa = -2$, respectively. The spin density $|n\mathbf{s}|$ is shown in the background, with $n_{\textrm{max}} = 5.1 \times 10^{-4}N_{\textrm{a}}/a_{r}^3$. The field of view is 15.5$\times$15.5 $\mu$m$^2$.}
\end{figure}

\begin{figure}[t]	
\includegraphics[width=0.95\columnwidth]{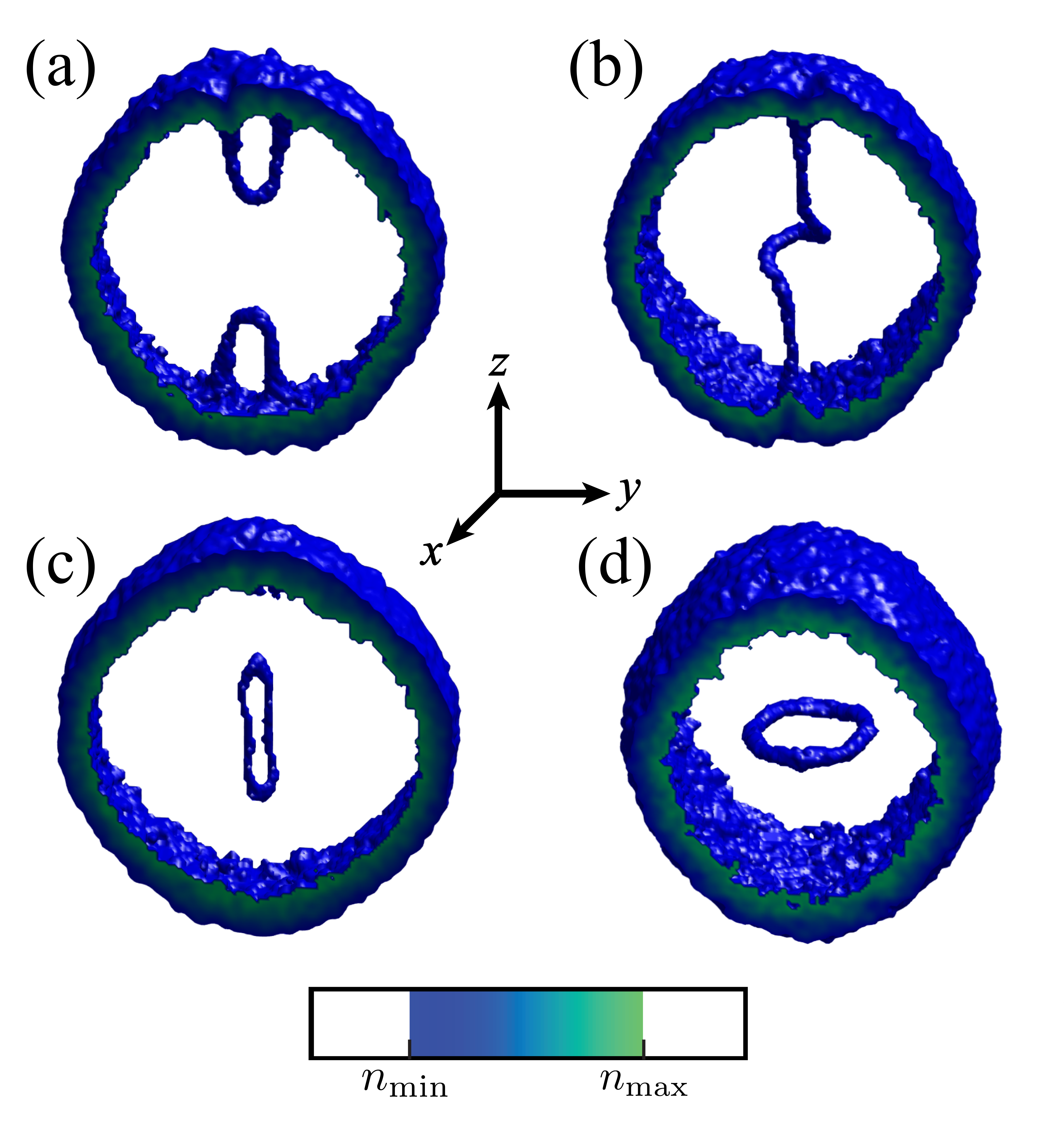}
\caption{\label{fig:5}{Decay of Dirac dipoles.} Isosurfaces of the spin density after 500 ms of decay of a Dirac dipole (a)~without additional vortices, (b)~initially pierced by a $\kappa=-1$ vortex, and (c,d)~pierced by a $\kappa = -2$ vortex. In cases (a)--(c) the two magnetic field zeros are kept in different hemispheres as in Fig.~\ref{fig:1}(c), whereas in (d) they coincide at the origin as in Fig.~\ref{fig:1}(d). The density range shown is [$n_{\textrm{min}}$, $n_{\textrm{max}}$] = [1.0, 5.1]$\times 10^{-4}N_{\textrm{a}}/a_{r}^3$. The field of view is 15.5$\times$15.5 $\mu$m$^2$.}
\end{figure}

We choose the parameters of the simulation  to be close to those of the experiment of Ref.~\cite{Ray2014.Nat505.657}. The number of trapped atoms is $N_{\textrm{a}} =  \int \Psi^{\dagger}(\mathbf{r},t)\Psi(\mathbf{r},t)  \mathrm{d}^3 r = 2.5 \times 10^5$, the optical trap frequency is $\omega = 2\pi\times 90$~Hz, and the material parameters are taken according to the natural values of $^{87}$Rb, namely, $a_0=5.387$~nm and $a_2=5.313$~nm~\cite{Kem2002.PRL88.093201}. 

We first find the ground state of the BEC in the uniform magnetic field shown in Fig.~\ref{fig:1}(b) using a successive overrelaxation method~\cite{Pre1994.book}. The subsequent monopole creation dynamics are simulated by solving the GP equation~\eqref{eq:GP} with the help of an operator-splitting method and fast Fourier transforms on graphical processing units. The computed area is $(24 \times a_r)^3$ with $a_r = \sqrt{\hbar/m\omega}$. The grid size is $200$ points per dimension and the time step is $2\times 10^{-4}/\omega$.

\section{Numerical results}\label{results}
We numerically simulate the proposed creation procedure by solving the dynamics according to the GP equation~\eqref{eq:GP} and by employing the parameters given above. {\note During the 80-ms adiabatic creation ramp, the offset field $\delta B$ is decreased linearly from 20 mG to 0.1~mG, which corresponds to the rate $\dot{B}_{\textrm{b}} = -0.25$~G/s.} This creation ramp places the magnetic field zeros $10 \%$ of the condensate radius away from the center of the condensate, as shown schematically in Fig.~\ref{fig:1}(c). Figure~\ref{fig:3} presents the results of the simulations analogous to the analytical results of Fig.~\ref{fig:2}. The corresponding spin fields of the BEC are shown in Fig.~\ref{fig:4}.

{Qualitatively, the GP simulations are observed to reproduce the Dirac dipole configurations of the analytical considerations in all three cases $\kappa = 0,-1,-2$. Quantitatively, however, we observe noticeable deviation from the analytical solution of Fig.~\ref{fig:2}, which we attribute to the vanishing magnetic field $\mathbf{B}_{\textrm{tot}}(\mathbf{r},t)$ between the field zeros as they approach the plane $z=0$. Consequently, the monopoles in Fig.~\ref{fig:3} come to a halt approximately halfway between the center and the periphery of the condensate even though the magnetic field zeros are brought closer to the center.}

The misalignment between the magnetic field direction $\hat{\mathbf{B}}_{\textrm{tot}}$ and the direction of the spin field $n\mathbf{s}$ is evident from Fig.~\ref{fig:4} for all the cases $\kappa=0$, $\kappa=-1$, and $\kappa=-2$: the field zeros are displaced from $z=0$ by $10 \%$ of the condensate radius, while the monopoles are located much further apart. {{Note also that the monopoles penetrate deeper into the condensate in the cases $\kappa=-1$ and $\kappa=-2$ than in the case $\kappa=0$. We attribute this difference to the existence of vortices already prior to the creation ramp when $\kappa \in \{-1,-2\}$, which results in vanishing condensate density along the paths of the magnetic field zeros.}} In all three cases, the spin density vanishes along the vortex cores. The monopole lag effect also adds artifacts in the case $\kappa=-2$ [Fig~\ref{fig:4}(d)]: both ends of the two-quantum vortex are attached to cone-shaped spin density depletions. We attribute this effect to the interplay between topological and energetic reasons: On the one hand, the spin should rotate by $4\pi$ around the monopole locations. On the other hand, the misalignment of the magnetic field and the spin field increases the Zeeman energy. This energy is lowered by forming the observed spin density depletions.

Next, we investigate the long-time evolution of the imprinted dipole configurations kept in the field of the form shown in Fig.~\ref{fig:1}(c). As shown in Fig.~\ref{fig:5}, the vortices undergo a relaxation process as expected from energetic considerations. In particular, the doubly quantized vortices in the case $\kappa=0$ split into pairs of singly quantized vortices, which can also be considered as single vortices changing their sign at the monopole locations. Due to this relaxation, the repulsion between vortices also decreases, and the monopoles shift towards each other, as shown in Fig.~\ref{fig:5}(a). For the case $\kappa=-1$ {[}Fig.~\ref{fig:5}(b){]}, the kinetic energy of the vortex with a reversed sign bends the vortex core near the condensate center. A simple classical analogy is a piece of rubber band whose ends are rotated in one direction and the center in the opposite direction. 

Whereas the creation ramp results in qualitatively similar monopole--antimonopole configurations in both final states of Figs.~\ref{fig:1}(c) and~\ref{fig:1}(d), the decay products for the case $\kappa=-2$ heavily depend on the final configuration of the magnetic field. On the one hand, if the magnetic field zeros do not coincide, as in Fig.~\ref{fig:1}(c), the monopole and the antimonopole are pinned to different hemispheres of the condensate, and the doubly quantized vortex between them splits into a pair of single-quantum vortices~{[}Fig.~\ref{fig:5}(c){]}. On the other hand, if the magnetic field zeros coincide at $(\rho,z)=(0,0)$, as in Fig.~\ref{fig:1}(d), the dipole evolves into an axisymmetric polar-core vortex-ring configuration~[Fig.~\ref{fig:5}(d)].

\section{Conclusion}\label{conclusions}
In summary, we have introduced and modeled a robust method to create Dirac monopole--antimonopole pairs in spinor BECs. Furthermore, we studied the creation of such dipoles when the initial state hosts a $\kappa$-quantum vortex and numerically simulated the creation and subsequent decay for $\kappa \in \{ 0,-1,-2\}$. We showed that the detailed structure of the created Dirac dipole, as well as their eventual decay products, change substantially with the value of $\kappa$.

{\note The proposed scheme is similar to the method of creation of Dirac monopoles~\cite{Pie2009.PRL103.030401} used in Ref.~\cite{Ray2014.Nat505.657} and can be realized in the existing experimental setups after a modification of the control coils. However, the strength of the magnetic field required to control the spins may exceed experimentally achievable values of resistive coils, and alternative ways to generate the magnetic field need to be sought. We suggest superconducting magnets as one possible solution.}

Dirac monopoles and antimonopoles in the natural magnetic field have been proposed to appear in pairs in a number of publications~\cite{MUSSET1983333,PhysRevD.29.1524,PINFOLD1993407, PhysRevLett.85.5292,TheH1Collaboration2005}. If natural magnetic monopoles were found, all the cases discussed in this paper should also be regarded as possibilities for the wave function of a charged scalar particle interacting with a pair of such monopoles.\\

\begin{acknowledgments}
The authors gratefully acknowledge funding support from the Academy of Finland through its Centres of Excellence Program (Project No.~312300) and Grant No.~308632, the Magnus Ehrnrooth Foundation, and the Technology Industries of Finland Centennial Foundation. This project has received funding from the European Research Council (ERC) under the European Union's Horizon 2020 research and innovation programme under grant agreement No 681311 (QUESS). CSC - IT Center for Science Ltd. (Project No. ay2090) and the Aalto Science-IT project are acknowledged for providing computational resources. The authors thank Ian Spielman and David Hall for inspiring discussions. 
\end{acknowledgments}

\bibliographystyle{apsrev4-1}
\bibliography{reflist,Dirac_dipole_footnotes}

\begin{thebibliography}{41}
\bibitem [{not({\natexlab{a}})}]{note_asterisk}%
  \BibitemOpen
  \bibinfo {note} {Please note that the asterisk does not denote complex
  conjugation.}\BibitemShut {Stop}%
\end{thebibliography}
\end{document}